\documentclass[10pt,twocolumn]{article}
\usepackage[letterpaper,margin=.76in,columnsep=.24in]{geometry}
\usepackage[T1]{fontenc}
\usepackage{lmodern}
\usepackage{amsmath,booktabs,tabularx,microtype,xcolor}
\usepackage[numbers,sort&compress]{natbib}
\PassOptionsToPackage{hyphens}{url}
\usepackage[hidelinks]{hyperref}
\usepackage{graphicx}
\usepackage{balance}
\definecolor{navy}{HTML}{184464}
\newcommand{\sys}{\textsc{Rbwa}}

\title{Release-Bound Withdrawal Authorization\\for Bitcoin Custody}
\author{Jian Cui\textsuperscript{1}\quad
Rui Zhu\textsuperscript{2}\quad
Haixu Tang\textsuperscript{3}\\[4pt]
\small\textsuperscript{1}University of Illinois Urbana-Champaign\\
\small\textsuperscript{2}Yale University\\
\small\textsuperscript{3}Indiana University Bloomington}
\date{November 2026}
\hypersetup{
  pdftitle={Release-Bound Withdrawal Authorization for Bitcoin Custody},
  pdfauthor={Jian Cui, Rui Zhu, Haixu Tang},
  pdfsubject={Bitcoin custody, withdrawal authorization, and fault recovery}
}
\begin{document}
\maketitle
\newcommand{\RaceCount}{20}
\newcommand{\ForgedSats}{70,000}
\newcommand{\LateSats}{49,000}
\newcommand{\OldPending}{50,000}
\newcommand{\FixedSats}{147,000}
\newcommand{\RollingSats}{98,000}
\newcommand{\DrainedSats}{258,000}
\newcommand{\HotBound}{260,000}
\newcommand{\ReserveSats}{939,000}
\newcommand{\BatchSats}{100,000}
\newcommand{\EffectCases}{14}
\newcommand{\Trials}{20}
\newcommand{\ModelStates}{54,048}
\newcommand{\ModelEdges}{145,440}
\newcommand{\LatencyN}{30}
\newcommand{\BaseMedian}{549.64}
\newcommand{\BaseTail}{608.85}
\newcommand{\RbMedian}{533.86}
\newcommand{\RbTail}{605.51}
\newcommand{\BaseRepeatMedian}{540.31}
\newcommand{\BaseRepeatTail}{592.43}
\newcommand{\RbRepeatMedian}{547.59}
\newcommand{\RbRepeatTail}{584.19}

\newcommand{\FaultRaces}{48}
\newcommand{\FaultRelease}{26}
\newcommand{\FaultPause}{22}
\newcommand{\FaultHistories}{50}
\newcommand{\FaultOperations}{392}
\newcommand{\LegalCases}{40}
\newcommand{\LegalMined}{4}
\newcommand{\NegativeCases}{128}
\newcommand{\IssuerFixedAMedian}{43.16}
\newcommand{\IssuerFixedATail}{89.43}
\newcommand{\IndependentFixedAMedian}{43.91}
\newcommand{\IndependentFixedATail}{77.48}
\newcommand{\IssuerRollingAMedian}{45.67}
\newcommand{\IssuerRollingATail}{75.97}
\newcommand{\IndependentRollingAMedian}{40.95}
\newcommand{\IndependentRollingATail}{78.63}
\newcommand{\ProofFixedAMedian}{-0.49}
\newcommand{\ProofFixedALow}{-9.49}
\newcommand{\ProofFixedAHigh}{1.22}
\newcommand{\ProofRollingAMedian}{-4.11}
\newcommand{\ProofRollingALow}{-13.11}
\newcommand{\ProofRollingAHigh}{2.29}
\newcommand{\WindowIssuerAMedian}{-0.96}
\newcommand{\WindowIssuerALow}{-11.37}
\newcommand{\WindowIssuerAHigh}{11.86}
\newcommand{\WindowIndependentAMedian}{0.16}
\newcommand{\WindowIndependentALow}{-10.33}
\newcommand{\WindowIndependentAHigh}{13.52}
\newcommand{\IssuerFixedBMedian}{42.41}
\newcommand{\IssuerFixedBTail}{219.47}
\newcommand{\IndependentFixedBMedian}{45.95}
\newcommand{\IndependentFixedBTail}{162.87}
\newcommand{\IssuerRollingBMedian}{48.19}
\newcommand{\IssuerRollingBTail}{90.23}
\newcommand{\IndependentRollingBMedian}{44.39}
\newcommand{\IndependentRollingBTail}{77.46}
\newcommand{\ProofFixedBMedian}{-1.22}
\newcommand{\ProofFixedBLow}{-10.17}
\newcommand{\ProofFixedBHigh}{10.12}
\newcommand{\ProofRollingBMedian}{-1.31}
\newcommand{\ProofRollingBLow}{-8.60}
\newcommand{\ProofRollingBHigh}{7.60}
\newcommand{\WindowIssuerBMedian}{-1.47}
\newcommand{\WindowIssuerBLow}{-3.95}
\newcommand{\WindowIssuerBHigh}{12.27}
\newcommand{\WindowIndependentBMedian}{3.11}
\newcommand{\WindowIndependentBLow}{-6.36}
\newcommand{\WindowIndependentBHigh}{7.22}

\begin{abstract}
A compromised cryptocurrency withdrawal backend can request signatures
on unauthorized payments. Previously signed transactions may remain
spendable after pause or become spendable again after a reorganization.
Release-Bound Withdrawal Authorization (\sys) orders permits against
pause in a replicated journal and restores exposure when its closing
evidence disappears from Bitcoin's active chain. An optional route
delays large payouts and permits independent cold recovery. Direct
payments confirm in one block on regtest, and staged payments confirm
in three or seven at delays of two or six. After rollback, a replayed
recovery can win or lose to an already exported mature payout.
Authenticated receipts can revoke one signing lease in a shared wallet.
In matched synthetic schedules, probes reduce permits through
compromised leases from 20 to 9 out of 33, with one false fence.
Two held-out HTTP workloads find equal fixed-threshold precision for
boosted audit and probe-augmented detectors. The audit fields already
expose the probe authentication outcome. Live tests verify an exact
maintenance exception and add a file marker as a distinct source.
The marker fences a compromised lease that RPC-only audit misses.
Audit with the same file-read evidence also fences a reader who
bypasses the probe. The results tie the benefit of active probing to
which evidence the operator can obtain.
\end{abstract}

\section{Introduction}
A Bitcoin node can verify that the required keys signed a transaction, while
an exchange must decide whether a customer actually authorized that
transaction. An attacker who controls a withdrawal backend can submit a
validly structured but unauthorized payment to an honest wallet signer. In its
preliminary account of the October 29, 2026 incident, Binance reported that
attackers reached its wallet signing infrastructure outside the normal
withdrawal workflow and that 18,270 BTC left a hot wallet~\cite{binance2026}.
The public account motivates our threat model. It does not disclose enough of
the signing architecture to support a claim that a particular protocol would
have prevented the incident.

Intent-Bound Withdrawal Authorization (IWAP) binds approval to decoded
transaction effects, reserves budget before signing, and requires separate
administration for higher-risk actions~\cite{iwap}. This is a useful starting
point. The published study uses a normalized admission monitor. A single
trusted intent issuer can still authenticate false instructions if compromised.
The specification calls for a current-policy check before release, but leaves
open how to order pause against an in-flight release when signers and policy
replicas fail independently. Its fixed-epoch budget also permits successive
settled withdrawals to draw on adjacent windows.

We present Release-Bound Withdrawal Authorization (\sys), a prototype for
a narrowly specified Bitcoin withdrawal profile. The design places the
irreversible decision to allow signing in the same replicated order as pause.
A prior decision remains charged even if its response is lost. The design
also requires independently verifiable consent and balance evidence, uses
rolling and unresolved-exposure limits, and separates the keys and funds of
the reserve from the online wallet.

\begin{enumerate}
\item We specify a release journal that orders pause, retry and settlement,
reopens closed liabilities after a chain reorganization, and supports
a separate cumulative-limit mode (Sections~\ref{sec:design} and
\ref{sec:properties}).
\item We implement the protocol over a three-member etcd cluster and native
Bitcoin SegWit transactions, including a mandatory policy co-signature and
an independently controlled reserve. A configurable large-batch route
binds delayed payout and recovery to the same permit. A separate lease
profile orders source-evidenced revocation against release
(Sections~\ref{sec:staged-design},~\ref{sec:lease-design}, and~\ref{sec:impl}).
\item We reproduce the public IWAP artifact and compare journal modes on
real Bitcoin spending and forced rollbacks. We test durable recovery
under faults, compare both outcomes of a rolled-back staged recovery, and
distinguish credential scope from file ownership and
mount isolation in independent local signing roles (Section~\ref{sec:eval}).
\end{enumerate}

\section{Background and threat model}
Bitcoin transactions consume previous outputs and create new outputs. A
withdrawal can include multiple recipients and a change output. A signer must
verify the source outputs, each recipient, the destination of change, and the
fee derived from the difference between input and output value. Native SegWit
separates witness data from the transaction identifier, and SegWit
\texttt{SIGHASH\_ALL} commits to all outputs and the input set for the
supported profile~\cite{bip141,bip143}. These properties let the policy
service bind approval to an exact unsigned transaction and verify the signed
template.

The primary adversary controls the withdrawal backend and its credentials.
It can construct arbitrary proposals, replay requests, schedule concurrent
operations, and present valid test workload evidence for an approved build.
We additionally consider compromise of the intent-issuer key, of the ordinary
hot-wallet key, and of both keys controlling the hot wallet. The last case
tests asset isolation after off-chain policy enforcement has been lost. These
are distinct experiments. The attacker can withhold a signature it already
obtained. Processes may stop or lose quorum access and later restart.

For authorization security we trust the customer's confirmation key and its
display and enrollment channel, the independent ledger, and a mandatory
policy signer. The experimental customer device and ledger use separate test
keys. Real production isolation would also require independent administrators,
recovery credentials, and deployment identities. For ordering we trust a
crash-fault-tolerant log, an authentic cluster identity, and the integrity of
its persisted state. A malicious quorum, administrative snapshot rollback,
and clock manipulation outside the stated clock assumption are outside the
claim. Tracked accounting assumes that the signer receives a useful
active-chain view from a trusted Core node. A sampled observation
does not establish instantaneous finality. We force rollbacks that
remove previously confirmed evidence. Retained accounting does not
recycle capacity based on such evidence.

\section{Design}\label{sec:design}
\begin{figure}[t]
\centering
\setlength{\fboxsep}{5pt}
\fbox{\parbox{.91\columnwidth}{\small
\textcolor{navy}{\textbf{1. Confirm and reserve}}\quad
Customer, ledger, and issuer authenticate the same intent.\\[3pt]
\textcolor{navy}{\textbf{2. Check actual effects}}\quad
The signer decodes the unsigned transaction and verifies proof freshness.\\[3pt]
\textcolor{navy}{\textbf{3. Order release and pause}}\quad
The wallet journal commits an exact-transaction release permit.\\[3pt]
\textcolor{navy}{\textbf{4. Sign and reconcile}}\quad
The trusted publisher retains signed bytes and reports the chain outcome.
}}
\caption{The critical order lives in the wallet journal. The chain
policy determines when an earlier permit's charge can be released.}
\label{fig:design}
\end{figure}

\paragraph{Independent withdrawal evidence.}
A customer intent contains a unique identifier, customer account, wallet,
network, destination, amount, fee ceiling, and expiration. The issuer signs a
canonical record. The customer signs the same fields in a separate domain.
The ledger verifies the customer proof, atomically places a balance hold for
principal plus the fee ceiling, and authenticates the corresponding record.
A replay of the same record can recover its receipt without a second debit.
A conflicting record with the same identifier fails. The policy signer
verifies the separate issuer, customer, and ledger proofs and a short-lived
workload record bound to the wallet, intent identifiers, and unsigned bytes.
The workload record can establish admissibility of an approved deployment.
Customer approval remains a separate prerequisite.

\paragraph{Verification of Bitcoin effects.}
The signer decodes the proposed bytes with Bitcoin Core. It obtains each
previous output from its trusted node, verifies that it belongs to the
selected wallet, and rejects duplicate inputs and proposals that already
contain signatures. It compares the multiset of external outputs to the
authorized recipients and amounts, permits at most one registered change
output, and checks the actual fee against the sum of signed fee ceilings.
Fee charges are assigned deterministically across the batch while staying
within each ceiling. The permit binds the exact bytes by hash, including
version, locktime, and sequences. Core signs with \texttt{SIGHASH\_ALL}; a
subsequent template check compares the non-witness transaction identifier
before and after signing. Changing the bytes after authorization requires a
new decision. The prototype rejects fee replacements under an already
reserved intent. It does not implement an automated replacement workflow.

\paragraph{Ordering issue, pause, and retry.}
A journal shard represents one independently capped wallet lane. It stores a
policy, epoch, pause bit, and an ordered event stream. \texttt{Reserve}
consumes the intent identifiers and charges the budget against the digest.
At this point the signer still lacks a release permit.
\texttt{Issue} rechecks pause, epoch, expiration, and the reserved digest,
then commits a release permit using the same compare-and-swap mechanism that
commits \texttt{Pause}. If a concurrent pause changes the state first, the
release CAS loses and a new read observes the pause. Resuming increments the
epoch, preventing a pre-pause reservation from becoming newly issuable
under the restored policy. Expiration is checked against service time
sampled before the CAS and applies to new permits. A reservation may be canceled while its state
still prevents issuance. An issued permit may be recovered on retry, with
the original identifier and log position.

A successful CAS defines the time of authorization. Signature creation,
delivery, and broadcast can happen later. The signer can recover an issued
permit after a lost response and produce the same transaction after pause.
This behavior accounts for the possibility that signature material already
escaped. The old permit stays charged. The design promises an order for
new permits and a bound on old liabilities.

\paragraph{Publication and reconciliation.}
The signing side stores completed bytes in a local durable outbox before its
publisher sends them to Bitcoin Core. A withdrawal backend can receive the
transaction identifier. The prototype's raw signing API is internal to the
test process. Duplicate publication checks the same transaction identifier.
A missing transaction or an RPC timeout leaves the charge in place.
After the configured confirmation depth, a trusted reconciliation step
either records settlement of the authorized transaction or verifies a
confirmed transaction sharing one of its inputs. The latter establishes a
conflict at that chain tip. The publisher checks terminal journal state
before publishing again. We use a reversible closure in the tracked mode.

\paragraph{Limits and the reserve.}
Let $O(t)$ be principal plus fee charges on unresolved reservations and
issued permits. Let $S_W(t)$ be charges on withdrawals settled within the
preceding $W$ seconds. At each admission a wallet checks
$O(t)+S_W(t)+x\le B$ and $O(t)+x\le U$, for charge $x$, spending limit $B$,
and pending limit $U$. The window uses the journal's trusted time, whose
observed value is clamped to its last committed value. Moving an unresolved
charge to settlement keeps it in the same budget initially. Later it ages
out of the rolling window. This prevents adjacent-window resets from
doubling near-boundary capacity. It permits cumulative spending over a
longer compromise, so it complements asset separation and refill control.
The implementation measures charges in satoshis for one wallet lane. It
does not share a global budget across lanes.

The hot wallet spends under a two-key native P2WSH policy. An ordinary hot
key and an independently controlled policy key must both sign. The reserve
has a separate Bitcoin key and independent approval lane. Refill approvals
bind an allowed hot-wallet destination and amount. With both hot keys
compromised, the attacker can bypass the hot journal. Provided reserve keys
and refill authority are intact, the available hot principal over a chosen
attack window is bounded by its opening spendable balance plus all new
external inflows, including authorized refills. A refill thus increases
exposure even when it follows policy. The reserve's input cannot be spent
with the hot keys.

\subsection{Reopening exposure after a reorganization}\label{sec:recovery-design}
The original journal permanently released the pending charge when a
confirmed conflict appeared. The conflicting transaction could later
leave the active chain, making an earlier signature usable again.
Keeping the original permit and the evidence for its closure gives the
journal enough information to reverse that accounting decision.

In the tracked mode, a closure stores the transaction identifier,
block hash, height and minimum depth. For invalidation, the signer
also verifies that the confirmed transaction spends an input of the
original transaction. The minimum depth applies at closure. A rollback
of descendant blocks leaves a closure in place when its anchor remains
on the active chain. This separates confirmation depth from evidence
that the original input is still spent on the observed branch.

Before admission, release, publication, or recovery, the signer reads
the journal and samples the trusted Core tip. It walks immutable
parent headers from that exact hash to the stored anchor heights,
then checks that the reported tip still matches. The walk does not
mix blocks from different branches when the active tip briefly
changes and returns. A missing anchor reopens the associated permit and
restores its full charge. One etcd compare-and-swap commits all
reopenings from that observation and sets a chain hold. This can expose
a pending total greater than the configured limit. The hold prevents
additional authorized output while the total is above the limit.
The signer rechecks around local signing and before and after
publication. A missing or moving pre-effect observation refuses
the operation. A post-send observation can restore accounting
after Core already accepted publication. The check is also
available explicitly to the operator.

The journal revision fences a closure check against concurrent journal
mutations. If another signer has changed the journal, the observation
must be refreshed. The chain hold persists across ordinary pause and
resume commands. An operator may recover the lane after reconciling
all reopened permits against the active branch, observing the chain
again, and checking both caps. A separate manual pause remains in
effect. An unavailable acknowledgement can leave the outcome unknown,
so a new worker reads the journal rather than subtracting a second
charge or assuming that the hold failed.

An existing lane can activate the new policy without creating an
empty budget. The operator first quiesces the legacy workers and
pauses the lane. One journal transaction switches the mode, reopens
old terminal records with their original permits, and sets the hold.
The operator opens that same lane under the new policy. A legacy
handle rejects the new policy. Recovery of old records requires the
original transaction data and fresh evidence. Lost legacy data keeps
the lane held.

The retained mode makes a different availability choice. It preserves
every issued permit's charge after settlement or confirmed conflict,
even when the active-branch evidence is intact. A reservation that never
issued can still be canceled. As a result, no later withdrawal can
reuse the capacity of an issued signature. This imposes a finite
cumulative issuance limit per lane. The capacity cost is present on a
stable chain. An upgrade can inherit prior issuance above that limit
and remain held. With arbitrarily deep reorganization and ordinary Bitcoin
signatures, confirmation count alone gives no irreversible point at
which to recycle that capacity.

\paragraph{Two commit points.}
Core can change branches after a sampled observation and etcd can
commit against an observation that was valid when read. No atomic
operation spans the journal and Bitcoin's chain or network. The
post-commit check can withhold local signing bytes when it sees a
rollback. A rollback inside a subsequent publication call can still
let Core accept the transaction before the journal reopens exposure. Also,
a party that already holds the full signatures can submit them without
this publisher. The tracked mode restores accounting on detection and
fences further authority. The retained mode supplies a separate
conservative bound on what the lane has already authorized.

\subsection{Staged payouts for large batches}\label{sec:staged-design}
An opt-in wallet lane stages a batch when its total authorized principal
meets a persisted threshold. The policy also identifies the staging
descriptor, payout and recovery roles, cold destination, and operator's
Ed25519 verification key. The signer validates the ordinary withdrawal
proposal and binds exact staging, payout, and recovery templates to one
reservation. The existing hot and policy keys co-sign the parent, which
funds a P2WSH staging output and sends any change to the registered change
address. The payout spends the staged output to the approved recipients.
The independent recovery key spends the same output, with the coordinator
checking the configured cold destination. Customer and workload evidence
must be fresh when the journal first permits the bundle. The authorized
child can mature after the original evidence expires.

The staging script follows the delayed-spend and immediate-recovery pattern
of Bitcoin vaults~\cite{bitcoinVaults}. The payout branch enforces a
block-based relative delay of $D$ using BIP~68 and
\texttt{CHECKSEQUENCEVERIFY}~\cite{bip68,bip112}. With payout sequence $D$
and stage confirmation at height $h$, the payout is first eligible for
block $h+D$. The recovery
branch has no relative delay. Both spends consume the same outpoint, so a
confirmed recovery excludes the payout on that branch. A reorganization
can remove that recovery or change the stage's inclusion height. The
current branch then determines which spends remain possible.

The journal orders payout authorization against recovery requests using
compare-and-swap. The operator separately authenticates the exact signed
recovery transaction. The coordinator verifies the cold witness and its
full-output signature commitment before the journal durably records it.
A recovery request
that commits first fences new payout authorization. The trusted publisher
checks the lane again near the send. An externally held payout can still
compete after maturity. A held lane can rebroadcast the recorded recovery
bytes, including after the recovery's confirmation anchor is lost. Stage
confirmation leaves the full charge pending. The reservation closes on a
confirmed exact payout, requested recovery, or proof that a different
transaction consumed a stage funding input. Recovery conservatively moves
the full original charge into the rolling budget under tracked accounting.

The caller's authorized fee total must cover the stage and the applicable
child. Staging adds a parent transaction and a relative wait, and the
matched comparisons keep the authorized fee total equal. The script
enforces the delay and signing branches. Destination enforcement remains a
signer policy, and recovery relies on a live key and an opportunity to
confirm before a competing payout. Independently authorized sub-threshold
batches can still use the direct path. The integrated regtest fixture uses
distinct payout and recovery wallets in one Core process.

\subsection{Revoking an individual signing lease}\label{sec:lease-design}
A lane can require a short-lived, independently issued withdrawal
lease. The lease identifies the wallet, lane, workload, generation,
holder key, and validity interval. Its holder signs a message for the
exact proposed transaction digest and requested operation. Reservation
and release use separate message domains. Customer, ledger, and
workload approval checks still apply. The journal binds the reservation
and permit to the lease's identifier. Multiple leases use the same
intent index and budget. A new lease does not increase either.

A decoy endpoint can issue a short-lived challenge for a live lease.
The endpoint has no signing or spending capability. Its receipt contains
the signed challenge, the holder's answer, and a separate sensor
signature attesting that the answer was observed. The signer checks all
three signatures and the original lease certificate. Opening a copied
URL or answering the challenge with a different lease key supplies
insufficient evidence. The receipt identifies use of a particular
credential. It conveys no assurance about who controlled the credential
or their intent. An audit source with a distinct signing key can also
provide a verified receipt for an exact lease.

Both sources use one journal transition. The lease fence competes with
release in the same compare-and-swap order. A fence that commits first
cancels the target's unissued reservations and releases their charges.
Their intent indexes prevent reassignment. Already issued permits keep
their charges. A replay returns the same fence, and reusing an event
nonce for another lease fails. A new generation has a distinct ID.
Retiring the old generation requires explicit revocation or expiry.
The trusted signing and publication paths recheck the fence. Separately
authorized cold recovery can proceed. A complete signature already held
elsewhere can still be broadcast under Bitcoin's own rules.
Lease expiry limits acquisition of a permit and signing through the
request path. Previously authorized publication and staged continuation
check the explicit fence, which can remain relevant after lease expiry.

An optional response policy treats a probe alone as advisory. It forwards
the probe when a fresh, independently signed weak audit hint identifies
the same lease. The hint has its own message domain and cannot authorize
revocation by itself. A strong audit receipt can still act immediately.
This policy runs before the shared executor. Its enforcement depends on
keeping the direct executor behind that trusted boundary.

An independently signed maintenance ticket can make one authenticated
probe advisory even when it has a corroborating hint. The ticket binds
the lease, namespace, lane, wallet, and exact challenge nonce. It
must be issued between challenge creation and probe observation, and
remain valid at the decision. It carries no signing authority. Strong
audit bypasses the exception. Invalid or cross-challenge tickets leave
the ordinary decision in force. The change authority is separate from
the holder, issuer, sensor, and audit keys.

An optional second witness provides information absent from ordinary
request authentication. A provisioner places a unique, lease-scoped
secret marker in a protected artifact. A holder who obtains the marker
can MAC the exact probe challenge, including its nonce and scope. The
trusted response verifies the original holder proof and the marker MAC
before promoting an otherwise advisory probe. The marker has no spend
authority. Exact maintenance remains advisory, and strong audit keeps
its usual path. The MAC establishes possession of the marker. Attribution
to a particular source depends on how the marker was provisioned and
whether it could be copied through another route.

\section{Conditional properties}\label{sec:properties}
\paragraph{Consent under issuer compromise.}
Assume the policy signer executes the checks, the customer confirmation key
is intact, and the ledger authenticates only a hold that it created for the
same verified intent. A release for an unauthorized recipient or amount
would require an accepted customer proof for a different message, a mismatch
between decoded and signed bytes, or an alternate spending path. The first
violates the proof assumption, and the remaining cases violate exact-effects
or signer-coverage assumptions. The argument does not protect a customer
who confirms the attacker's address or whose confirmation key and device
are compromised.

\paragraph{Fencing and recovery.}
Assume linearizable transactions and trusted signer execution. For a
particular lane, the log orders a racing issue attempt and pause.
If pause commits first, later issue attempts fail while the lane remains
paused. An earlier state read cannot be committed after the pause with
the stale revision. After resume, an old reservation's epoch fails
validation. An issued record may be returned multiple times for the same
digest without producing another authorization event. This is a property
of logical issue, and the accounting includes an older permit that
materializes after pause. Failures of quorum access can suspend liveness.

\paragraph{Exposure accounting.}
A reservation increases both counters only after testing the current caps.
Issuing or retrying a permit preserves the charge. A cancel releases the
charge only if no permit can have been issued. Settlement moves the charge
from pending into the rolling spend window in the legacy and tracked
modes. A confirmed conflict releases the pending charge in those modes.
In the tracked mode, the atomic reopen transition adds the charge back
and holds the lane, including when the corrected total already exceeds
the cap. The invariant covers states after successful reconciliation.
Between observations the actual chain can move first. For retained
accounting, every issued permit remains charged and issuance can only
follow a reservation within the caps. Thus a change in the chain cannot
free a retained charge or increase the sum of charges for issued permits
under one policy and lane. This bound assumes intact journal state and
a signing path that requires its permits.

\section{Implementation}\label{sec:impl}
The Python prototype uses Ed25519 for authorization proofs, SQLite for the
test account ledger and local signed outbox, and etcd v3.6.7 for the wallet
journal. Each etcd transaction updates the wallet's state and appends an
audit event together. The simple implementation rewrites a JSON wallet
snapshot on each transition. This is convenient for the bounded study and
leaves log compaction and larger admission volumes open. Anchor
verification caches immutable headers, but a cold walk can grow
with the distance to the oldest relevant anchor. The trusted
signing path queries Bitcoin Core v32.1 through the loopback RPC interface.
It uses raw native SegWit transactions rather than PSBT interchange. Core
validates the real script witnesses, accepts signed bytes into a local
mempool, and mines blocks on regtest.

The main Bitcoin fixture keeps the hot, policy, and reserve keys in
separate wallet descriptors in one Core process. Additional experiments
use four Core processes, first under one macOS user and then in
Linux containers with distinct non-root identities and private mounts.
The isolated signing fixture transfers blocks and raw signatures
through a trusted local test controller. It exercises Bitcoin spending
and role isolation separately from the etcd recovery workload.
A production deployment also needs a customer approval interface and
account-hold reconciliation. The reference ledger tests durable placement
and idempotence of holds.

\section{Evaluation}\label{sec:eval}
\paragraph{Setup and comparators.}
The experiments use Bitcoin Core 32.1 and etcd 3.6.7 on one Apple Silicon
Mac17,8 with 18 logical CPUs and 24 GiB RAM. The journal has three
loopback members. Except for the isolation experiment, the same Core
fixture is used for both treatments. The main comparisons retain native
Bitcoin bytes, wallet keys, the exact-effects checker, etcd-backed issue,
and the outbox. They vary proof requirements, accounting windows, or the
pending cap. The snapshot ablation removes the atomic issue decision.
We report synthetic inputs, forced schedules, and local timing as such.
All signing and confirmation claims use Bitcoin Core on regtest.

\paragraph{IWAP reproduction.}
On the pinned public commit we reproduced all 1,200 constructed
rejections and 3,000 valid admissions. In 20 concurrent trials, five of
64 requests reserved the 100-unit cap. All non-timing summary fields and
all 1,200 CSV rows matched the archive. The archived summary and macros
give different timing medians. We use neither for performance comparisons.
The upstream bounded model permits a reserved intent to enter the signed
state after revocation. The paper's prose separately asks signers to
verify current policy~\cite{iwap,iwapArtifact}. Our transactional issue
design addresses the gap between a policy observation and its use.

\subsection{Separating the policy controls}
The matched experiment sends the same signed proofs and unsigned
withdrawals through a $2\times2$ design with issuer-only or independent
proofs and fixed or rolling budgets. The four lanes use separate
counters but the same preimages and Bitcoin wallet. Whenever several
lanes approve the same payment, one submits it and all applicable
lanes reconcile the same confirmed transaction. The scope is one
controlled schedule per cell.

\begin{table}[t]
\centering\small
\begin{tabularx}{\columnwidth}{@{}l X r r@{}}
\toprule
Proofs & Window & Forged & Boundary \\
\midrule
Issuer only & Fixed & 70,000 & 147,000 \\
Independent & Fixed & 0 & 147,000 \\
Issuer only & Rolling & 70,000 & 98,000 \\
Independent & Rolling & 0 & 98,000 \\
\bottomrule
\end{tabularx}
\caption{Matched controls, in satoshis of principal. Forged measures
Core-acceptable principal under the issuer-key compromise. Boundary
measures confirmed honest principal across the budget boundary.}
\label{tab:factorial}
\end{table}

Table~\ref{tab:factorial} isolates each effect. Independent proof
checking rejects the forged 70,000-sat request under either window.
Rolling accounting excludes the third 49,000-sat payment under either
proof policy. This is a safety and availability tradeoff at the tested
boundary. In a separate comparison with rolling accounting on both
sides, a 60,000-sat pending cap blocks a second 50,000-sat request while
the first signature is unresolved. Turning off just the pending cap
permits both, reaching the 100,000-sat rolling limit.

\subsection{Concurrent release and failure}
The original Bitcoin experiment forces pause to win \RaceCount{}
times and confirms zero later permits. A snapshot-based ablation uses
a checked but outdated state and produces a Core-acceptable
\LateSats-sat transaction after pause. A previously committed permit
stays charged and can still result in a usable signature. This distinction
motivates the protocol's accounting rule.

The expanded experiment uses independent clients on each fresh lane.
A seeded pre-CAS barrier lets both clients read before either
commits. It schedules 16 release-biased, 16 pause-biased, and 16
near-tie cases. All \FaultRaces{} observed invocations overlap.
Release wins \FaultRelease{} and pause wins \FaultPause{}. After resume,
all attempts using an unused reservation from the previous epoch fail,
and the newly reserved controls succeed. These counts establish coverage
of both commit orders. The forced schedule does not estimate how often
either would win in deployment.

A loopback HTTP bridge witnesses an etcd commit before withholding its
response. One trial closes the connection. A second kills the separate
client worker before it can receive the response. In both trials we then
stop the elected etcd leader, allow a majority to elect its successor,
and recover the original permit with a new client. Each log contains one
release event. The earlier charge remains outstanding after pause and
window rollover. The offline checker replays \FaultHistories{}
authoritative histories with \FaultOperations{} operation intervals and
finds no violation of the checked invariants. It rejects corrupted
sequence and digest controls. A separate ablation shows that even a
linearizable read can precede a pause that commits before its intended
output. The checker rejects the output marker because it has no
corresponding issue event. This ablation exercises journal semantics;
the Bitcoin signing consequence is measured separately in the original
regtest experiment.

The separate finite explorer visits \ModelStates{} states and
\ModelEdges{} directed transitions with no modeled violation in the
authoritative configuration. It supplies a short counterexample for
snapshot-only signing. Neither a bounded exploration nor the offline
history checker establishes general implementation refinement.

\subsection{Transaction semantics}
We construct seeded cases with a separate Bitcoin wire reader and a
predeclared table of funding outpoints, recipient scripts, amounts and
fees. The oracle compares the unsigned output multiset and then the
signed non-witness template. It is separate from the Core-backed policy
checker. Legal cases include two to four inputs, shuffled payments,
repeated identical outputs, no change and fees exactly at the authorized
ceiling. All \LegalCases{} legal cases sign and pass Core validation.
We broadcast and confirm \LegalMined{} of them. The other candidates
are checked separately and may share inputs.

All \NegativeCases{} mutations across 16 families fail at the expected
stage with no added journal event or signed outbox record. These include
recipient substitution, an amount swap that preserves total value,
aggregating identical payments, duplicate or mixed-wallet inputs,
unregistered change, transplanted proof or evidence, and mutation after
reservation. Most adversarial cases receive valid fresh workload
evidence, avoiding an unrelated rejection. Seven direct-sign
diagnostics on output mutants are all Core-acceptable. Those controls
show that the authorization policy, rather than Bitcoin's validity
rules, accounts for their rejection.

\subsection{Key and RPC separation}
We provision native P2WSH spends across four loopback Core processes
with separate data directories and RPC credentials. Two fresh trials
synchronize validated blocks to the hot, policy, and reserve nodes.
With the scoped hot credential, cross-role RPC calls receive HTTP 401.
Calls to load another wallet by name or path, export descriptors, or
make a backup receive HTTP 403 on the hot endpoint. Core rejects the
hot-only witness and confirms the 49,000-sat two-signature payment.
The two online roles together can drain a different 199,000-sat
payment, while Core rejects their attempt to spend the separate reserve
input. The reserve role confirms a 60,000-sat refill, leaving 939,000
sat including the refill's fee deduction.

The stronger negative control uses the unrestricted hot administrator.
With the reserve process stopped and both nodes running as one OS user,
that administrator loads the reserve wallet by absolute path and
produces a Core-acceptable reserve signature in both trials. We
withhold it. Distinct processes and cookies alone therefore do not
contain this administrator in the tested deployment.

In two follow-up trials we reuse the same fixture and change only the
hot process launcher. A local macOS filesystem policy denies that
process access to the reserve data directory. The unrestricted hot
credential then receives Core error -4 on the same wallet-load attempt
in both trials. The legitimate hot payment and authorized refill still
confirm. This is a test of the specific path on one host. It leaves
alternate copies, other role directories, and operating-system
administrator compromise outside the tested boundary.

\paragraph{Separate Linux role domains.}
We repeat the signing controls in two fresh Docker fixtures inside
one Linux ARM64 VM. Each of the four Core roles has a different
non-root UID/GID, its own RPC admin cookie, and its own mode-0700
wallet volume. Role containers run without
network access, a host filesystem bind, Docker socket, Linux
capabilities, or permission escalation, and use a read-only root
filesystem. A trusted test controller transfers Core RPC requests
and blocks through container-local execution. We model compromise of
the unrestricted hot RPC credential and code running as the hot UID.

In both trials, the hot admin cannot load the reserve wallet by
name or absolute path. The reserve path and cookie are absent from
the hot role, and the hot credential fails reserve RPC authentication.
Mounting the reserve volume into a diagnostic container under the
wrong UID produces an actual permission denial. Repeating with the
reserve UID and the same mount reads the wallet and cookie, loads
the stopped wallet and produces a Core-acceptable 938,000-sat reserve
spend, which we withhold. This positive attack control distinguishes
credential checks from storage ownership. Each isolated trial still
confirms the 49,000-sat joint payment and controller-approved 60,000-sat
refill. The two online keys can spend a separate hot output for
199,000 sat but cannot sign the reserve input. The reserve retains
939,000 sat after the refill and its fee.

\subsection{Confirmation-depth counterexample}
The original fixture withholds a signed 49,000-sat withdrawal and
settles a competing transaction of the same input. The expanded
experiment tests configured depths of two, three, and six blocks,
each under a stable chain, a tip-only rollback, and a rollback that
removes the conflict. Each arm rejects reconciliation at depth $d-1$
and accepts it at $d$, after which the signer authorizes a new
withdrawal from a disjoint input. The journal carries 50,000 sat,
including fees, against a 60,000-sat pending limit.

In all six stable and tip-only scenarios, Core rejects the earlier
signature. In all three conflict-removal scenarios, Core accepts both
signatures and mines both withdrawals into the same alternative block.
The block pays 98,000 sat of principal. Immediately before publication
the two signatures represent 100,000 sat including fees, while the
journal retains a charge of 50,000 sat. The trusted publisher rejects
the invalidated record in all nine cases. A separate holder of the old
signature can still submit it directly. This counterexample violates the
confirmation assumption deliberately. It bounds the claim even at a
greater fixed depth, without estimating the frequency of natural
reorganizations. These runs use the legacy journal and motivate the
recovery experiment below.

\subsection{Reversible closure and failure recovery}\label{sec:recovery-eval}
We cross the three accounting modes with confirmation depths two,
three, and six and four controlled schedules, for 36 real Core
cases. Each begins with one valid withheld signature and a second
disjoint input. Each withdrawal charges 50,000 sat. The rolling
budget is 100,000 sat and the pending cap is 60,000 sat.
Settlement attempts at one below the selected depth fail. The
matched modes differ in whether and how a closure releases the
old charge.

\begin{table}[t]
\centering\small
\begin{tabularx}{\columnwidth}{@{}p{.35\columnwidth} X X X@{}}
\toprule
Schedule & Legacy & Tracked & Retained \\
\midrule
Stable or tip-only & New grant & New grant & Cap blocks \\
Conflict removed before new grant & 50k charged & 50k, held & 50k, held \\
Conflict removed after new attempt & 50k charged & 100k, held & 50k, held \\
\bottomrule
\end{tabularx}
\caption{Pending charge in satoshis and lane decision at the first
post-rollback observation. Each schedule is tested at three depths.
The 50k legacy value charges the newer permit only. Retained
accounting blocks the second grant before or after rollback.}
\label{tab:recovery}
\end{table}

Table~\ref{tab:recovery} separates rollback timing from the accounting
policy. A surviving conflict anchor preserves tracked availability,
even when a tip-only rollback temporarily lowers its depth. Removing
the anchor before another grant makes tracked mode restore the first
50,000-sat charge and stop new permits. When rollback happens after
the new permit, tracked mode accounts for 100,000 sat and holds the
lane. Legacy still charges only 50,000 sat. Old and new complete
signatures already held externally can still mine together. Recovery
reconciles reopened permits on the new chain, and a separate operator
action resumes the lane. Retained mode prevents the second grant
even in the stable cases. It counts the first 50,000 sat throughout.

\paragraph{Core and journal faults.}
An injected observation failure denies new signing and leaves the
journal and outbox without a new grant or signed bytes. A later coherent
observation can proceed. In a second diagnostic we remove the
conflict after the observation but before the issue CAS. The journal
can still commit that permit. The post-commit check restores exposure
and withholds the new signature. The issued permit remains charged.
In a distinct publication fault we inject the rollback inside the
final Core send. Core accepts the transaction before the post-call
check restores the 100,000-sat total. Both disjoint withdrawals then mine
98,000 sat of principal. This provides a concrete boundary to any
claim of atomicity with Bitcoin's network.

\paragraph{Mixed-branch observations.}
We force Core from branch A to a higher-work branch B and back to A
inside one RPC interval. A mutable height lookup returns B's closure
anchor even though both boundary tip reads return A. Repeating the
actual excursion during the immutable-header walk still identifies
A's ancestors. The signer restores 50,000 sat and holds the lane.
In a separate control a tip change persists into the final read.
The signer rejects with no new permit or outbox bytes, and a fresh
coherent observation permits the retry.

\paragraph{Lost acknowledgement of the hold.}
Two further experiments use real three-member etcd with synthetic
chain observations. A loopback proxy drops the acknowledgement
after the atomic reopening and hold for two records, and we kill the elected
leader. A fresh client finds exactly one hold event in each mode.
The tracked lane moves from 25,000 to 90,000 charged units despite
the 80,000-unit cap. The retained lane stays at 65,000 and denies
an extra 25,000-unit reservation. A stale retry fails and a fresh
repeat adds neither a charge nor an event. Ordinary resume cannot
clear the hold. After all orphans resolve and capacity is available,
explicit recovery leaves the separate manual pause intact.

\paragraph{Activation and repeated rollback.}
We also upgrade a paused legacy lane after its conflict has already
left the active chain. The upgrade restores 50,000 sat under the
original permit, preserves its evidence, and fences the old policy
handle. We mine the old withdrawal, settle it on the new branch
and clear the chain hold while the manual pause stays active. Rolling
back this newly confirmed settlement restores 50,000 sat again.
Repeating the observation leaves the same journal sequence and does
not add another hold event.

\subsection{Large-batch latency and recovery}\label{sec:staged-evaluation}
We compare the tracked direct baseline with the same implementation
configured to stage batches at 100,000 sat of principal. We test
99,999 and 100,000 sat at two configured delays, with two repetitions
in reversed arm order, for 16 runs. Both arms use a 3,000-sat total
fee for successful payouts. We count blocks from immediately before
the initial send to first confirmation of the customer payment. The
fixture mines an eligible transaction promptly. Reconciliation
requires two confirmations of the terminal transaction.

\begin{table}[t]
\centering\small
\begin{tabular}{@{}lrrr@{}}
\toprule
Principal (sat) & Delay & Direct & Enabled \\
\midrule
99,999 & 2 & 1 & 1 \\
99,999 & 6 & 1 & 1 \\
100,000 & 2 & 1 & 3 \\
100,000 & 6 & 1 & 7 \\
\bottomrule
\end{tabular}
\caption{Blocks to first customer confirmation on regtest, with two
matched repetitions per row. Enabled sub-threshold payments add zero
blocks. Staging adds one transaction and the configured delay.}
\label{tab:staged-latency}
\end{table}

All repetitions agree (Table~\ref{tab:staged-latency}). A two-recipient
40,000-plus-60,000-sat batch also stages. Confirming only the parent
leaves the full charge pending. A separate small payment confirms
while a large payment waits on the same lane. A later recovery rollback
holds a new small permit on that lane. Negative controls reject a changed
recipient, inadequate fees, template substitutions, an unconfigured
journal handle, and invalid recovery credentials. The coordinator also
rejects a corrupted witness and a valid signature that omits the
commitment to the recovery outputs.

\paragraph{A recovery reorganization has two outcomes.}
For the six-block policy, the fixture exports a complete payout and
then records an independent recovery request. The initial recovery
confirms and the journal closes the permit. Removing its closing block
with the stage still present restores 103,000 sat and holds the lane.
The worker reannounces only the original recovery in each of two
controlled replacement-branch schedules. When the miner includes the
recovery in the next block, the funds return to reserve and the payout
conflicts. When the miner omits it until payout maturity and includes
the old payout, the payout transfers 100,000 sat. The journal settles
the actual winner. Neither branch creates a second withdrawal permit.

\paragraph{Parent rollback and ambiguous send.}
In two controls the same parent transaction reconfirms on a later
branch. Core rejects the leaked child until the delay from the new
inclusion height elapses. Without a recovery request the payout first
confirms two blocks after restaging. With the request, the worker sends
recovery and the miner includes it in the next block. In a separate
fault Core accepts recovery and the RPC acknowledgement is lost. A
fresh worker reannounces the exact recovery with the same txid and
creates neither a new signature nor a new permit.

An additional funding conflict first closes the permit because it
spends the stage input. Rolling it back restores the original charge.
The worker then relays the cached parent and recorded recovery during
the hold, and both confirm. These forced schedules establish outcomes
conditional on inclusion. Public-network latency and rollback frequency
remain outside the measurements.

\subsection{Scoped revocation and source coverage}\label{sec:lease-eval}
We enroll two signing lease holders in the same wallet. A valid probe
revokes only A. The journal cancels A's unissued 50,000-sat reservation,
reducing outstanding exposure from 90,000 to 40,000 sat. B confirms two
39,000-sat payments, using released headroom for the second reservation.
Copied challenges, cross-holder proofs, and a signed reissue of the same
nonce for B fail. Replaying an old receipt leaves a newly issued lease
usable. Neither a fresh lease nor a rotation duplicates the wallet
budget or allows another lease to reuse the canceled intent.

We force both commit orders with two journal clients. Revocation first
prevents any A permit or signing outbox entry. Release first leaves
50,000 sat charged and the exported signed payment remains acceptable
to Core. B confirms a payment in both orders. A loopback proxy also
drops the HTTP reply after a committed audit revocation. We stop the
elected etcd member, reopen the lane, and find one fence at the same
sequence. A remains blocked and B confirms. The staged route retains
the independent cold recovery path when the initiating lease is revoked.

\paragraph{Matched event schedules.}
We replay 13 predeclared synthetic schedules in three arms, for 39
runs. The audit rule reacts immediately to a trusted audit event with
a corroborated lease and abstains when attribution is missing. The
other arms use probe evidence alone or accept either source. They use
the same signer, journal transition, transaction bytes and ordinary
approval proofs per schedule. Each of the 11 compromised schedules
includes three otherwise policy-valid A requests. We measure permits
through the compromised credential. The fixture does not label the
underlying payments as nonconsensual and withholds all signed payments.

\begin{table}[t]
\centering\small
\begin{tabularx}{\columnwidth}{@{}Xrrr@{}}
\toprule
Outcome & Audit & Probe & Either \\
\midrule
Compromised permits out of 33 & 20 & 15 & 9 \\
Fence before first attempt out of 11 & 3 & 5 & 7 \\
Correctly fenced compromised traces & 7 & 8 & 10 \\
False fences & 0 & 1 & 1 \\
Healthy A requests blocked & 0 & 3 & 3 \\
Healthy B permits out of 13 & 13 & 13 & 13 \\
\bottomrule
\end{tabularx}
\caption{All arms share the same lease revocation executor. Counts
describe the selected schedules, with misses retained in denominators.}
\label{tab:lease-comparison}
\end{table}

Adding probe evidence lowers issued principal through compromised
leases from 980,000 to 441,000 sat (Table~\ref{tab:lease-comparison}).
In two schedules the audit stream lacks sufficient attribution and a
valid probe identifies A before release. Audit wins when it comes first
or the attacker bypasses the decoy. When both are late, all three
requests precede the fence. The valid probe of a healthy holder falsely
fences A and blocks three legitimate requests in the probe arms.
The combined rule increases coverage on these schedules and incurs a
false action absent in the audit-only arm. Logical event times are set
by the experiment. These results give no estimate of production alert
latency, trap engagement, or false-alarm rates.

\paragraph{Corroborated response.}
We freeze a separate suite of nine synthetic schedules and replay each
under strong audit alone, either-source response, and the corroborating
policy. The latter two admit 147,000 sat through compromised leases,
compared with 245,000 sat for strong audit. In the early joint case,
the corroborating policy commits its fence at trace offset 6, compared
with 25 for strong audit, and admits zero requests instead of two.
An isolated benign authenticated probe remains advisory. A wrong
lease, lane, key or stale hint fails to corroborate. False fences fall
from six with either source to one with corroboration. Strong audit
has zero false fences. The jointly benign probe and hint still cause
one false fence, and the weak-only attack still obtains all three
permits. All 27 B requests are authorized. The response policy reduces
collateral interruption relative to immediately acting on probes in
these schedules, while the comparison with strong audit remains a
trade-off between early containment and false actions.

\subsection{Held-out revocation precision}\label{sec:precision-eval}
The scripted ordering experiment controls when the decoy is reached.
We therefore run a separate loopback service
and record 8,046 completed HTTP operations from 896 generated episodes.
The server verifies holder and origin proofs, actual probe responses,
and exact maintenance tickets. The driver records the compromised
lease separately. Calibration and test blocks use different seeds.
Each test block contains a fixed mix of benign and adversarial
workloads, including direct bypass, benign origin changes, and
credential misuse from the expected host. The 512 test episodes cover
192 compromised A leases and 832 healthy A or B leases.

We give all arms ordinary audit fields, including the operation,
authentication, attested origin, and maintenance status. Boosted audit
and augmented arms use the same causal prefix features and model
settings. The augmented arms add either all verified probes or verified
probes without authorized maintenance. An audit rule supplies a
nonlearned comparator. Thresholds come from disjoint calibration blocks
at a 2\% empirical false-action budget. Each lease contributes one
decision before its first sign request. The score does not see episode
IDs, families, labels, or elapsed time. We estimate differences by
paired bootstrap of whole test blocks and distinguish constructed
attack prevalence from field precision~\cite{arpSecurityEvaluation}.

\begin{table}[t]
\centering\small
\begin{tabularx}{\columnwidth}{@{}Xrrrr@{}}
\toprule
Model & TP & FP & Precision & AP \\
\midrule
Audit rule & 0 & 0 & -- & .152 \\
Boosted audit & 27 & 12 & .692 & .516 \\
+ verified probes & 27 & 12 & .692 & .516 \\
+ maintenance-aware probes & 27 & 12 & .692 & .516 \\
\bottomrule
\end{tabularx}
\caption{Pre-signing decisions on 1,024 held-out leases. Positive labels
identify 192 compromised leases. Each learned arm has 14.1\% recall
and 1.44\% false-action rate at its frozen threshold.}
\label{tab:heldout-precision}
\end{table}

The three learned arms tie at the calibrated operating point
(Table~\ref{tab:heldout-precision}). Their decisions match
boosted audit on every lease. The maintenance-aware AP difference is
0.00023, with a paired 95\% interval from $-0.00062$ to $0.00135$.
For the recorded probe operations, the holder-verification bit agrees
with ordinary authentication in every event. It supplies no independent
classification signal in this fixture. We also verify 48 opposite-label
pairs with equal observed pre-signing features. Any deterministic rule
on those inputs must fence both or neither in each pair. With at most
16 false actions (2\% of 832 healthy leases, rounded down), even an
otherwise perfect rule could fence at most 160 of the 192 compromised
leases on this test mixture.

An independent seed generates another 512 test episodes and 4,535
HTTP operations. We retain the original calibration data and frozen
scorer. All learned arms yield 31 true and 15 false actions on the
replication, with 16.1\% recall and 67.4\% precision. An exploratory
sweep over six calibration false-action budgets from zero to 10\%
gives matching true and false counts across learned arms on both
test sets. The replication varies generated workloads and keys
within the fixture, without testing a new deployment population.

\paragraph{Maintenance and fault controls.}
Eight cases on Bitcoin Core and three-member etcd exercise actual
response and payment. An exact maintenance ticket leaves its probe
advisory. A later explicit audit still revokes A. Wrong, stale, forged,
and cross-challenge tickets cannot suppress the response. In the
wrong-challenge case we lose the HTTP reply after commit, replace the
leader, and recover the original fence without a duplicate. B confirms
a payment in all eight cases. Benign operation with a missing ticket
can still cause a false stop. Bit-identical maintenance evidence also
produces the same advisory decision when the driver stipulates that
the holder and ticket were stolen.

\paragraph{A witness beyond ordinary authentication.}
An adaptive comparison crosses seven cases with RPC-only audit,
artifact-backed probes, and informed file-read audit. A subprocess
reads the generated marker from a private-mode temporary file and
produces the challenge MAC. The benign unscheduled probe and the
marker-read attack have the same normalized RPC inputs. RPC-only
audit fences neither of two controlled attack families. The artifact
proof fences the attack that probes, but misses a reader who goes
straight to signing. Audit with the same trusted read observation
fences both. The five healthy A payments and seven B payments confirm
per arm. Wrong marker, lease, and nonce proofs fail. This result
isolates the benefit of an additional source where ordinary RPC audit
lacks coverage. The lab instruments a same-user subprocess. It
establishes neither exclusive access to a production marker nor an
advantage over audit with equivalent evidence.

\subsection{Local service time of the original signer}
We run two fresh clusters with 30 measured blocks each, following two
warmups. A block reuses the same real transaction, inputs, and prepared
evidence across all four profiles, with seeded random profile order.
One profile publishes after all four profiles finish in each block. We measure preparation
and issue plus wallet signing separately from Core's mempool check
and publication. This avoids masking authorization cost with the
roughly 410-ms median local publication path. It also holds workload
and Core state fixed within each timed block.

\begin{table}[t]
\centering\small
\begin{tabularx}{\columnwidth}{@{}Xrr@{}}
\toprule
Profile & Run A & Run B \\
\midrule
Issuer / fixed & \IssuerFixedAMedian{} / \IssuerFixedATail & \IssuerFixedBMedian{} / \IssuerFixedBTail \\
Independent / fixed & \IndependentFixedAMedian{} / \IndependentFixedATail & \IndependentFixedBMedian{} / \IndependentFixedBTail \\
Issuer / rolling & \IssuerRollingAMedian{} / \IssuerRollingATail & \IssuerRollingBMedian{} / \IssuerRollingBTail \\
Independent / rolling & \IndependentRollingAMedian{} / \IndependentRollingATail & \IndependentRollingBMedian{} / \IndependentRollingBTail \\
\bottomrule
\end{tabularx}
\caption{Local median and p95 in milliseconds for preparation plus issue
and wallet signing. Each profile has 30 measured observations per run.}
\label{tab:timing}
\end{table}

Table~\ref{tab:timing} shows variation at the tail on the same host.
For example, the paired median of adding independent proofs with
rolling accounting is \ProofRollingAMedian{} ms in run A and
\ProofRollingBMedian{} ms in run B. Descriptive 95\% block-bootstrap
intervals are [\ProofRollingALow{}, \ProofRollingAHigh{}] and
[\ProofRollingBLow{}, \ProofRollingBHigh{}] ms, respectively. We
resample adjacent four-block chunks with 2,000 draws. All four
proof and window contrasts cross zero in both runs. The data do
not resolve a local timing advantage or establish equivalence.
Preparation starts after customer, issuer and ledger proof
generation. The timing study does not measure confirmation
latency, concurrent throughput, or deployment-scale performance.

\section{Related Work}
\label{sec:related}

\paragraph{Signals for revocation.}
Canarytokens send notifications when decoy artifacts are accessed, and
cloud audit logs can identify service-account credentials used in some
requests~\cite{canaryTokens,auditCredential}. IWAP describes alerts as
a possible cause for pausing an affected workload~\cite{iwap}. Our
comparison feeds both evidence sources to one journal operation and
distinguishes evidence of credential use from correctness of a detection.

\paragraph{Transaction authorization.}
Complete mediation requires checking authority at each use of a protected resource~\cite{completeMediation}.
Article 5 of the EU strong customer authentication rules requires dynamic linking between payment authentication and the amount and payee~\cite{dynamicLinking}.
In digital-asset custody, Fireblocks documents a customer-configured approval callback before MPC signing, and BitGo exposes destination, amount, and velocity policies for withdrawals that involve its service~\cite{fireblocks,bitgo}.
PSBT standardizes how signers exchange unsigned transactions, input context, and partial signatures. SegWit and BIP~143 define witness serialization and the version~0 signature digest~\cite{bip174,bip141,bip143}.
These mechanisms provide the policy and transaction interfaces at which RBWA operates. RBWA checks the authenticated intent against the transaction at the signing boundary and gives the decision to issue the result an explicit order relative to a pause.

\paragraph{Intent-bound custody.}
IWAP specifies independent verification of customer intent, exact transaction effects, and workload evidence, together with durable, single-use budget reservations~\cite{iwap}.
Its specification directs each signer to check current policy before releasing a signature and retains charges under uncertainty. It also discusses replenishment and unresolved exposure across epochs.
Its executable reference monitor admits synthetic records in SQLite. In its separate bounded explorer, a reservation can reach the signed state after revocation because only the reservation transition checks the pause~\cite{iwapArtifact}.
RBWA addresses the interval between the last authorization check and the issue decision. It orders that decision and pause in the same replicated state, retains exposure for transactions that may have escaped, and tests the boundary with native Bitcoin transactions on regtest~\cite{bitcoinCore}.

\paragraph{Ordering and retries.}
Linearizability orders concurrent operations at points between invocation and response~\cite{linearizability}. Raft implements a replicated log, and etcd exposes linearizable key-value transactions while allowing an RPC outcome to remain unknown to a client after a timeout~\cite{raft,etcd}.
Chubby sequencers let a protected server reject operations from stale lock holders~\cite{chubby}. Zanzibar orders authorization checks with changes to access policy and protected content, and RIFL durably records RPC completion to disambiguate retries~\cite{zanzibar,rifl}.
RBWA uses replicated ordering for issue and pause and binds the outcome of each issue attempt to an exact transaction. An issued Bitcoin transaction can subsequently be submitted through a different broadcaster, so policy revocation alone cannot reclaim the corresponding exposure.

\paragraph{Containment after authorization.}
Bitcoin vaults use time-delayed spending and a separate recovery path to limit hot-wallet exposure~\cite{bitcoinVaults}. RBWA accounts for transactions authorized under ordinary Bitcoin spending rules, where invalidation depends on chain state and the confirmation assumption~\cite{chainReorg}.
The Binance initial incident report illustrates a different form of persistent authority. The exchange reported that Ethereum delegations enabled additional asset movement after ordinary withdrawals were suspended~\cite{binance2026,eip7702}.
That episode motivates accounting for authority already issued. The RBWA prototype evaluates Bitcoin withdrawals.

\section{Limitations and deployment}
The study uses generated keys and test funds on one physical host.
Three local processes exercise real etcd consensus and crash recovery.
The log guarantees assume crash faults, trusted signers and intact
persistent state. The Linux roles share one kernel, container engine
and trusted controller. Their experiment exercises Unix ownership,
mount isolation and RPC credentials. It does not test a compromised
Docker administrator, shared storage administrator, kernel exploit,
remote service transport, or a separate machine per role. The integrity
of authorization proofs depends on enrollment and recovery separation.
The prototype's wallet caps do not implement cross-wallet, provider-wide
budgets. A provider that needs an aggregate cap can preallocate disjoint
lane budgets or serialize a common cap, with different availability costs.

Bitcoin's confirmation model cannot ensure deterministic finality.
Tracked accounting corrects its state on a successful observation. An
idle lane can lag until the next explicit synchronization or signer
operation. An unavailable, stale or isolated node limits what can
be observed. Core and etcd have independent commit points, and the
tests show that a previously issued signature can still be published
after the last pre-publication check. Retained accounting avoids
recycling charges and pays with a finite cumulative budget. Ongoing
multi-wallet capacity management, automated fee replacement, signer
hardware isolation, and customer balance reconciliation remain
deployment work. The Ethereum delegation path in the incident motivates
restricting action types. Our implementation is Bitcoin-only and exercises
neither EIP-7702 nor a comparable general-purpose smart-contract policy.
The tests give controlled comparisons on the declared threat model.
They do not reconstruct Binance's systems.

The staged route requires an intact policy signer on the funding path
and a separate recovery operator. The prototype exposes an explicit
reconciliation call. The experiments supply the calls and control block
inclusion. They establish neither response time under network congestion
nor confirmation priority for recovery. Fixed fees can leave recovery
unable to replace an already relayed payout. A stage created by a single
large batch receives the delay. Separate approved smaller batches can
still use the direct path, subject to the wallet caps. The shared lane
hold can also delay small withdrawals during an observed rollback.

The lease profile trusts the issuer, sensor, and audit identities as
configured and uses synthetic credentials. The comparison prescribes
when each source sees an event. It can test the consequence of observing
an event before release, but cannot estimate how often a real attacker
will touch a decoy or how quickly a deployed alerting stack will respond.
The HTTP study executes cryptographic operations and captures their
times. Its workload families and attack labels remain constructed.
Its learned decisions are evaluated before a simulated sign boundary.
Separate live controls test the actual signer and revoke operation.
Neither result estimates error rates for a commercial audit service.

\section{Conclusion}
\sys{} orders signature permits against pause and reopens exposure
when a recorded closing block leaves the observed chain. This lets
the signer stop further grants and recover a previously lost charge.
Complete signatures remain usable outside that accounting boundary,
and a cumulative retained budget trades capacity for a stronger
bound. Tests with separate Linux identities and volumes show when
a hot role can and cannot reach reserve signing authority.
Large batches can use a delayed payout and a recorded cold recovery.
Replaying recovery after rollback preserves funds when recovery confirms
first on the new branch. A mature externally held payout can win instead.
An independently authenticated observation can revoke one workload
lease while other holders continue to use the shared wallet and budget.
On held-out local workloads, probe verification adds no precision when
ordinary audit already exposes the same authentication evidence.

\paragraph{Artifact and disclosure.}
The research artifact includes the protocol implementation, the
experiment runner and raw results, the bounded state model, and reproduction
records for IWAP. Codex assisted the research design, implementation, and
manuscript preparation, using an in-session AutoResearch cycle and
the scientific brainstorming workflow described
in~\cite{scientificAgentSkills}. The manuscript reports locally
executed results and explicit model assumptions.

\balance
\begingroup
\raggedright
\small
\setlength{\bibsep}{3pt}
\bibliographystyle{unsrtnat}
\bibliography{refs}
\endgroup
\end{document}